\documentclass[epsfig]{aa}
\usepackage{graphics}

\def\kms{km~s$^{-1}$}

\def\Msun{M$_{\odot}$}

\def\sch{Schwarzschild}
\def\ha{H$\alpha$}
\def\hb{H$\beta$}
\def\hg{H$\gamma$}
\def\hd{H$\delta$}

\begin{document}

\thesaurus{06(          
              08.16.4;  
              08.01.3;  
              08.12.1;  
              08.15.1;  
              08.22.3;  
              02.19.1)  
           }

\title{Envelope tomography of long-period variable stars\thanks{Based on 
observations made at Observatoire de Haute Provence, operated by the Centre 
National de la Recherche Scientifique, France}}
\subtitle{I. The Schwarzschild mechanism and the Balmer emission lines}

\author{Rodrigo Alvarez\inst{1}  
   \and Alain Jorissen\inst{1}
   \and Bertrand Plez\inst{2}
   \and Denis Gillet\inst{3}
   \and Andr\'e Fokin\inst{4}
}
\institute{IAA, 
           Universit\'e Libre de Bruxelles, 
           C.P.\,226, Bvd du Triomphe,
           B--1050 Bruxelles, Belgium
           (ralvarez,ajorisse@astro.ulb.ac.be)
\and
           GRAAL,
           Universit\'e Montpellier II,
           cc072,
           F--34095 Montpellier cedex 05, France
           (plez@graal.univ-montp2.fr)
\and
           Observatoire de Haute-Provence,
           F--04870 Saint-Michel l'Observatoire, France
           (gillet@obs-hp.fr)
\and
           Institute for Astronomy of the Russia Academy of Sciences,
           48 Pjatnitskaja,
           109017 Moscow, Russia
           (fokin@inasan.rssi.ru)
}

\offprints{R.\ Alvarez}
\mail{ralvarez@astro.ulb.ac.be}

\date{Received 6 April 2000/ Accepted 31 August 2000}

\titlerunning{Envelope tomography of LPV stars. I}
\authorrunning{R.\ Alvarez et al.}

\maketitle

\begin{abstract}

This paper is the first one in a series devoted to the study of the dynamics 
of the atmospheres of long-period variable (LPV) stars. Results from a 
two-month-long monitoring of the Mira variables RT\,Cyg and X\,Oph around 
maximum light with the ELODIE spectrograph at the Haute-Provence Observatory 
are presented. The monitoring covers phases 0.80 to 1.16 for RT\,Cyg and 
phases 0.83 to 1.04 for X\,Oph. 
The cross-correlation profile of the spectrum of RT\,Cyg with a K0\,III mask 
confirms that the absorption lines of RT\,Cyg in the optical domain appear 
double around maximum light. No line doubling was found in the optical 
spectrum of X\,Oph around maximum light, indicating that this feature is not 
common to all LPVs.

This paper also presents the application to RT\,Cyg of a new 
tomographic\footnote{The word {\it tomography} is used here in its 
etymological sense ({\it `display cuts'}), which differs somewhat from the 
broader sense in use  within the astronomical community (reconstruction of 
a structure using projections taken under different angles).} technique 
deriving the velocity field across the atmosphere by cross-correlating the 
optical spectrum with numerical masks constructed from synthetic spectra 
and probing layers of increasing depths. This technique reveals that both 
the temporal evolution of the line doubling, and its variation with depth 
in the atmosphere of RT\,Cyg, are consistent with the `\sch\ scenario'. This 
scenario relates the temporal evolution of the red and blue peaks of the 
double absorption lines to the progression of a shock wave in the atmosphere.

The temporal evolution of the Balmer \ha, \hb, \hg\ and \hd\ emission lines 
around maximum light is also presented for RT\,Cyg and X\,Oph. The velocity 
variations of \ha\ and of the absorption lines are discussed in the 
framework of two competing models for the formation of Balmer emission lines 
in LPV stars.
\end{abstract}

\section{Introduction}

Long-period variable stars (LPVs) are cool giant stars showing more or less 
periodic light variations with amplitudes of several magnitudes in the 
visual and with periods of several hundred days. Depending on their visual 
amplitudes and on the regularity of their variability cycles, they appear in
several flavours, namely Mira Ceti-type variables (Mira stars or Miras), 
Semi-Regular variables (of the SRa or SRb subtypes) or Irregular variables 
(of the Lb subtype). Mira and SR variables represent one of the latest 
stages in the evolution of stars with initial masses in the approximate 
range 1 to 9~\Msun.

It is known since long that the brightness variations of LPVs go along with 
spectral changes:
(i) hydrogen and some metallic lines turn from absorption to emission after 
minimum light (Merrill 1921), 
(ii) the velocity of the emission and absorption lines correlates with phase 
and excitation potential (Merrill 1923a; Adams 1941),
(iii) several absorption lines appear double around maximum light
(Adams 1941; Merrill \& Greenstein 1958; Maehara 1968).

In a review on red stars, Merrill (1955) was the first to suggest that the
bright emission lines sometimes appearing in LPVs may be explained by some 
kind of ``hot front'' moving outward. He further suggested that this running 
hot front may have the structure of a shock wave.

Although the shock wave scenario is nowadays quite widely accepted (see de 
la Reza 1986 and references therein), the lack of self-consistent pulsation 
models for LPV stars (due to the important role played by convection and the 
difficulty of modelling it) prevents from grasping exactly how, why and 
where emission and double absorption lines form. As a consequence, several 
(sometimes conflicting) theories aiming at explaining the spectral 
peculiarities of LPVs have appeared in the literature. Although the most 
popular models locate the formation of emission lines in the hot wake of the 
shock (e.g., Gillet 1988a), an alternative model proposed by Magnan \& de 
Laverny (1997) associates emission lines with purely radiative non-LTE 
processes independently of any shock wave. In this model, the double-peak 
nature of the emission lines is not connected in any way with the velocity 
field, contrary to the shock-wave model.

Concerning the absorption lines, their doubling around maximum light is most
easily seen in the rather clean near-infrared spectral domain (e.g., Gillet 
et al.\ 1985; Hinkle et al.\ 1997 and references therein). Long-term radial 
velocity measurements of the infrared rotation-vibration lines of CO have 
revealed well-defined variations with phase, regularly repeating from one 
cycle to the next and following a typical S-shaped curve (Hinkle et al.\ 
1997). These radial-velocity variations are generally believed to reflect 
the differential bulk motions occurring in the large and tenuous atmosphere 
of LPVs and associated with its pulsation.
Alternatively, the complex absorption profiles have also been explained 
without resorting to differential velocity fields: Gillet et al.\ (1985) 
have argued that the doubling of the metallic absorption lines results in 
fact from the development of an emission core in the line. This emission 
component is caused by the radiative release of the thermal energy dumped 
into the post-shock layer by the shock front. This model has the advantage 
of being able to account for the P~Cygni line profiles sometimes observed in 
LPV spectra (and appearing when the emission core becomes brighter than the 
local continuum).
A similar explanation in terms of the shock-induced temperature inversion 
has been proposed by Karp (1975) for the line doubling observed in Cepheid
variables. Detailed radiative-transfer calculations have shown that the
formation of double absorption lines does not necessarily require a velocity
gradient, since a temperature inversion is in principle sufficient.  

Recent studies have clearly validated theories accounting for the complex
absorption line profiles observed in several kinds of pulsating stars 
(RR\,Lyrae: Chadid \& Gillet 1996, Fokin \& Gillet 1997; $\beta$\,Cephei 
stars: Mathias et al.\ 1998; BL\,Herculis: Gillet et al.\ 1994, Fokin \& 
Gillet 1994; RV\,Tauri: Gillet et al.\ 1989a) in terms of velocity fields. 
The validity of the \sch\ scenario accounting for double absorption lines in 
terms of a velocity gradient in the photosphere remains however to be 
demonstrated for LPV stars. The present paper presents the first clear 
observational evidence thereof. 

\section{The \sch\ scenario}

\begin{figure}
  \resizebox{\hsize}{!}{\includegraphics{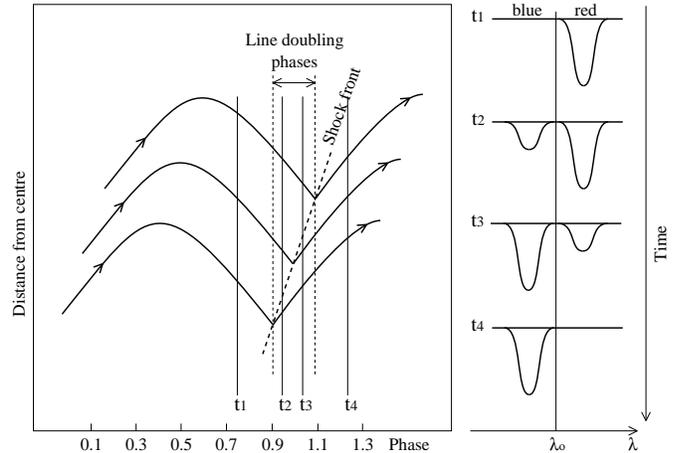}}
  \caption[]{The \sch\ scenario: temporal sequence followed by the intensity 
   of the red and blue components of absorption lines close to light 
   maximum, when the shock wave propagates through the photosphere, in the 
   absence of any complication due to radiative processes associated with 
   the shock wave}
   \label{Fig:Schwarzschild}
\end{figure}

A definitive way to conclude that double absorption lines are indeed the 
signature of a velocity gradient in the atmosphere rather than of a central
emission core is to check whether the double absorption-line profiles behave 
as predicted by the \sch\ scenario.
Although \sch\ (1952) proposed this scenario in relation with W\,Vir 
variables, it should be applicable to any kind of variable stars with a 
shock wave propagating through its photosphere. According to this scenario, 
the intensity of the blue and red components of a double line should follow 
(around maximum light) the temporal sequence illustrated in 
Fig.~\ref{Fig:Schwarzschild} (Note that this scenario implicitly assumes 
that the processes radiating away the thermal energy dumped into the 
post-shock layers by the shock front have a negligible impact on the line
profile). 
The observation of such a temporal sequence for the absorption line profiles
in Miras would definitely point towards the velocity stratification as the
origin of the double absorption lines, and moreover would offer a way to
directly probe the velocity field associated with the pulsation.

\section{The cross-correlation technique}

The extreme difficulty in testing whether the \sch\ scenario operates in LPV
stars comes from 
(i) the need for an intensive  mid-term (i.e., daily for about 2 months 
around maximum light) monitoring of LPV stars,
(ii) the crowded nature of the spectra of Mira stars, particularly in the 
optical domain. 
The cross-correlation technique provides a powerful tool to overcome the 
second difficulty. Indeed, the information relating to the line doubling 
(velocity shift and line shape) is in fact distributed among a large number 
of spectral lines, and can be summed up into an average profile, or more 
precisely into a cross-correlation profile, by a cross-correlation 
algorithm. If the correlation of the stellar spectrum with a specially 
designed mask involves many lines, it is possible to extract the relevant 
information from very crowded and/or low signal-to-noise spectra (e.g., 
Queloz 1995). Examples of the use of the cross-correlation technique (with 
the CORAVEL spectrovelocimeter) to derive line profiles may be found in 
Gillet et al.\ (1989a, 1990) for RV\,Tauri stars and in Udry et al.\ (1998) 
for Mira stars.

\section{Monitoring of the Mira variables RT\,Cyg and X\,Oph}
\label{Sect:monitoring}

\subsection{The ELODIE spectrovelocimeter}

A monitoring of two Mira-type stars, RT\,Cyg and X\,Oph, was performed with 
the fibre-fed echelle spectrograph ELODIE (Baranne et al.\ 1996) from 1999, 
August 3 to October 4, with an interruption between September, 3 and 
September, 16. Both stars were observed each night (weather-permitting), 
resulting in 32 spectra for both RT\,Cyg and X\,Oph. ELODIE is mounted on 
the 1.93-m telescope of the Observatoire de Haute Provence (France), and 
covers the full range from 3906~\AA\ to 6811~\AA\ in one exposure at a 
resolving power of 42\,000. Cross-correlation functions (CCF) with the 
default K0\,III mask have been computed directly in the (pixels,orders) 
space (i.e., without merging or rebinning the spectrum, in order not to lose 
accuracy) following the prescription of Baranne et al.\ (1996).

\subsection{RT\,Cyg}

\begin{figure*}
  \resizebox{\hsize}{22cm}{\rotatebox{90}{\includegraphics{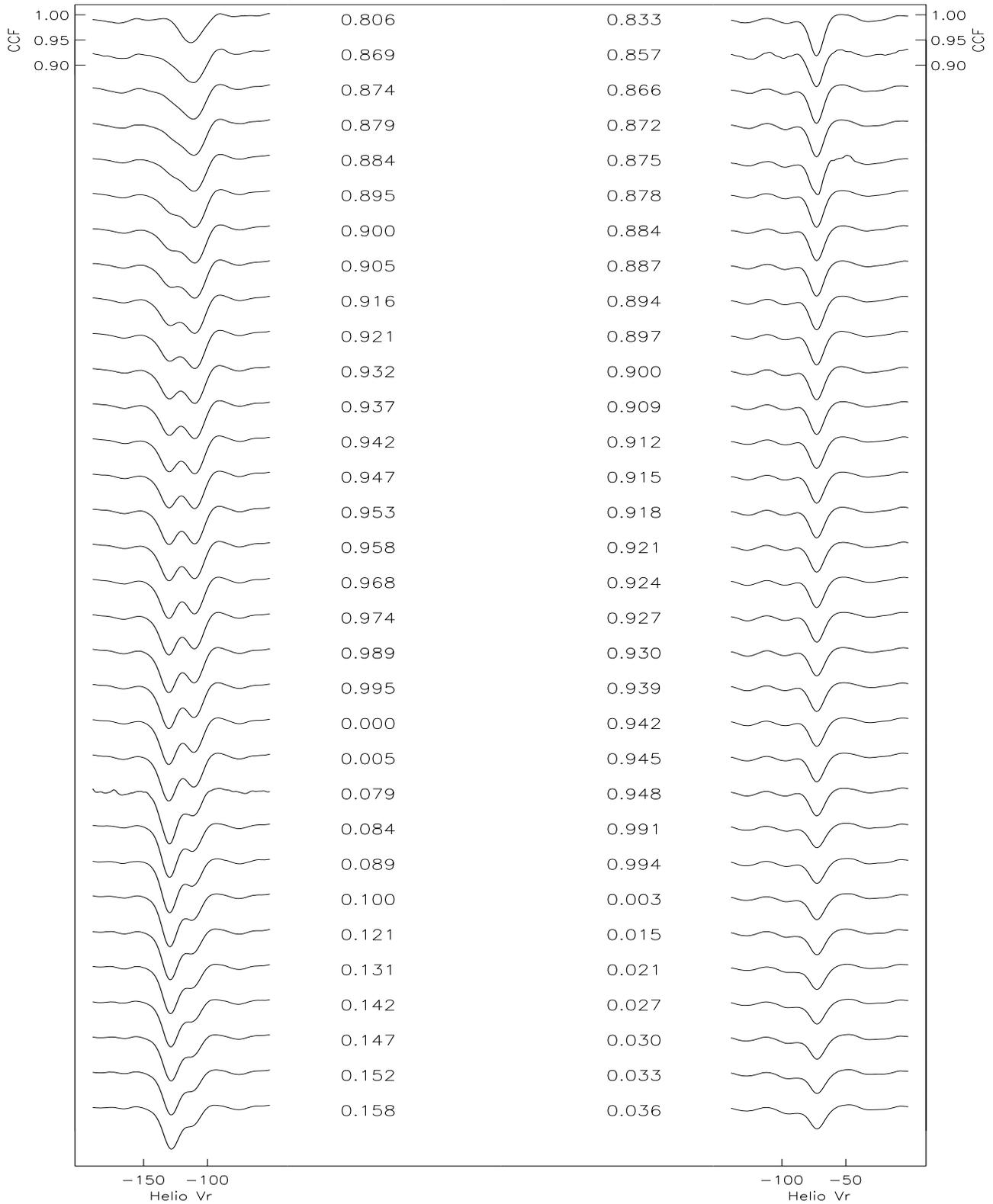}}}
  \caption[]{Sequence of cross-correlation profiles of RT\,Cyg (left 
  side) and X\,Oph (right side) obtained with the default K0\,III mask in 
  August-September, 1999. The labels beside each CCF denote the visual phase 
  based on the GCVS period and the AAVSO estimate of the epoch of maximum 
  light (1999, Sept.~1 for RT\,Cyg and 1999, Sept.~16 for X\,Oph; Mattei 
  1999, Observations from the AAVSO International Database, private 
  communication)}
  \label{Fig:suivi}
\end{figure*}

In Fig.~\ref{Fig:suivi}, the nightly sequence of CCF profiles obtained 
for RT\,Cyg is presented. 
RT\,Cyg is a M2e--M8.8e Mira star belonging to an old (extended/thick disk 
or halo) population (Alvarez et al.\ 1997), with a rather short period 
[190.28~d according to the {\it General Catalogue of Variable Stars (GCVS)}; 
Kholopov et al.\ 1988], typical of old populations, and a visual amplitude 
of 7~mag. Its maximum light was reached around 1999, September 1 according 
to the AAVSO light curve (Mattei 1999, Observations from the AAVSO 
International Database, private communication). 
Due to the uncertainties on the epoch of maximum light (2--3 days) and since 
the period is not constant from one cycle to the other, the transformation 
from Julian Dates to phases carries some degree of arbitrariness. In the 
present paper, phases were computed assuming that RT\,Cyg reached maximum 
light on 1999 September, 1 and adopting the GCVS period. It is worth noting 
that the light curve of RT\,Cyg shows a bump on the rising branch. By 
analogy with RR\,Lyr variables, this may be a signature of a particularly 
strong shock wave running through the atmosphere (Hill 1972; Fokin 1992; 
Bessell et al.\ 1996).

It can be seen in Fig.~\ref{Fig:suivi} that the sequence of CCFs 
follows the \sch\ scenario sketched in Fig.~\ref{Fig:Schwarzschild}: at 
phase $\sim$0.80, a single absorption peak is seen at $-$113~km~s$^{-1}$; 
then, coming closer to maximum light, the CCF becomes more and more 
asymmetric (phases 0.80--0.90), until the blue component becomes clearly 
visible around phase 0.90. The blue peak continues to strenghten and becomes 
even stronger than the red peak after phase 0.96. The intensity of the red 
component, that remained almost constant until this phase, then starts to 
fade away. A visual inspection of a few clean lines in the spectrum has 
confirmed that the doubling of the CCF is not an artefact of the method, but 
reflects the doubling of the spectral lines probed by the considered mask.

\begin{figure}
  \resizebox{\hsize}{!}{\includegraphics{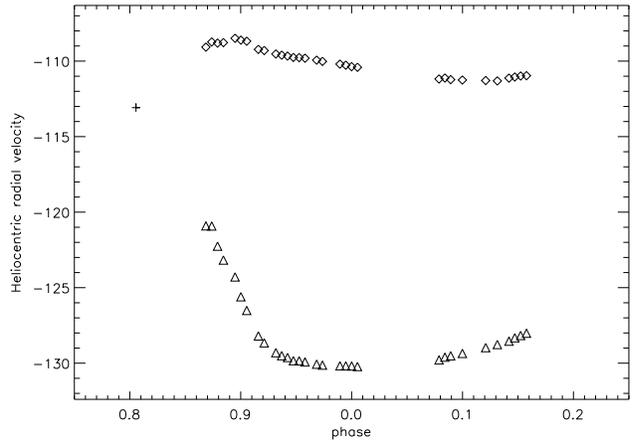}}
  \caption[]{Velocity variations of RT\,Cyg (cross: single component; 
   triangles: blue component; squares: red component}
  \label{Fig:Vr_rtcyg}
\end{figure}

The CCFs have been fitted by gaussian or double gaussian functions in order
to derive the corresponding radial velocity, with the results shown on
Fig.~\ref{Fig:Vr_rtcyg}.
The single absorption peak with $-$113~km~s$^{-1}$ seen at phase 0.80 splits 
up into a blue component with $-$121~km~s$^{-1}$ and a red component with 
$-$109~km~s$^{-1}$. The blue component becomes more and more blueshifted 
until its velocity reaches $-$130~km~s$^{-1}$ around phase 0.0. After 
maximum light, the blue shift of the blue component slowly decreases. The 
radial velocity of the red component appears to be much more stable than 
that of the blue component. The monitoring stopped at phase 0.158 (see 
Lebzelter et al.\ 1999 for radial velocities from infrared CO and Ti lines 
covering a more extensive fraction of the light cycle).

Unfortunately, the center-of-mass (COM) velocity is not known for that star, 
as there are no submm CO observations available\footnote{Based on a 
theoretical relationship  between the shock amplitude and the outward 
post-shock velocity, Willson et al.\ (1982) obtain $-119$~\kms\ for the COM
velocity if RT\,Cyg is pulsating in the fundamental mode, or $-123$~\kms\ 
for a first-overtone pulsation. Since these values are strongly 
model-dependent, they will not be used here.}.
The chances of detecting CO lines in RT\,Cyg are weak, though, since its 
IRAS 12~$\mu$m flux (6.4~Jy) is more than 10 times smaller than in stars 
with detected CO lines. The unknown COM velocity prevents us from putting 
constraints on the strength of the shock (see, however, the analysis of 
Willson et al.\ 1982). For the classical \sch\ scenario to hold true, the 
COM velocity needs to fall in between the red and blue peaks (as it is the 
case for all the Mira variables monitored in the IR by Hinkle et al.\ 1997, 
as well as for the theoretical estimate of the COM velocity of RT\,Cyg by 
Willson et al.\ 1982; see also Lebzelter et al.\ 1999). In that case, a new 
shock wave appears at the bottom of the photospheric layers as they are 
already moving downwards on the ballistic motion induced by the previous 
shock (Chadid \& Gillet 1996).
The blue peak observed in RT\,Cyg would then correspond to ascending matter 
in the envelope, driven by the coming shock wave. If on the contrary the COM 
velocity would not fall in between the red and blue peaks but both peaks are 
`redshifted' (a situation encountered at some phases in radial pulsators 
like RR\,Lyrae stars; see Chadid \& Gillet 1996), this situation corresponds 
to a receding shock (in an Eulerian rest frame; in Lagrangian coordinates, 
the shock is obviously propagating outwards). Such a shock is not strong 
enough to oppose the infalling atmosphere.

In principle, the velocity variations of the blue peak convey information 
about the propagation of the matter just behind the shock. The shock may 
either accelerate as it encounters less and less resistance in an atmosphere 
of decreasing density, or decelerate as part of its energy is dissipated by 
radiative losses during its upward motion (Gillet et al.\ 1985).
In RT\,Cyg (Fig.~\ref{Fig:Vr_rtcyg}), the two trends are observed in 
succession: the ascending layer which is first accelerating (phases 0.87 to 
0.98) decelerates after phase 0.08. The same trend may be observed for the 
layer where the red peak forms, which reacts with some delay to the motion 
of the underlying `blue' layer. At phase 0.9, when the outward motion of the 
blue layer is the fastest, the inward motion of the red layer is decelerated 
(the slope changes from positive to negative on Fig.~\ref{Fig:Vr_rtcyg}) as 
it is colliding with the rising shock. The outward acceleration stops at 
phase 0.0, and the red layer resumes its inward acceleration (on a possibly 
ballistic motion) as the force exerted by the blue layer has vanished (since 
the acceleration of the blue layer is now directed {\it inwards}). 

However, the interpretation presented above is very crude as it neglects 
possible complications coming from optical-depth or geometrical effects. The 
line formed in an expanding atmosphere will be asymmetric even in the 
absence of a velocity gradient (e.g., Karp 1975), simply because the 
projection of the expansion velocity on the line of sight varies across the 
stellar disk. In the presence of velocity gradients, the problem is further 
complicated by the fact that different lines of sight across the stellar 
disk probe different geometrical depths (i.e., layers with different 
velocities) at any given wavelength. Moreover, in the presence of shocks, 
the matter gets compressed, leading to a steep increase in the optical 
depth. If the location of this steep optical-depth rise changes with respect 
to the velocity field, velocity variations unrelated to the shock 
acceleration or deceleration may be recorded. Dynamic models including 
radiative transfer are therefore ultimately needed to correctly interpret 
the results presented on Fig.~\ref{Fig:Vr_rtcyg}.   

The maximum velocity discontinuity (observed around phase 0.98) amounts to
20~km~s$^{-1}$, translating into 26~km~s$^{-1}$ after correction for limb
darkening (Parsons 1972). Similar discontinuities are reported by Hinkle et 
al.\ (1997) for the two components of the double CO infrared lines observed 
in many Mira stars, which form however in different atmospheric layers than 
the optical lines sampled by the ELODIE mask.
Willson et al. (1982) have performed a detailed analysis of the velocity 
structure of the atmosphere of RT\,Cyg, based on photographic spectra 
covering the optical domain as does ELODIE. These authors claim to have 
observed {\it two} shocks, the deeper one having an amplitude of about 
30~\kms, and the upper one about 13~\kms\ (see their Fig.~5). Denoting by 
$v_A$ and $v_B$ the post- and pre-shock velocities of the deeper shock, and 
by $v_C$ and $v_D$ the post- and pre-shock velocities of the upper shock, 
they obtain $v_A = +10$~\kms, $v_B = -20$~\kms, $v_C \sim +6$~\kms\ and 
$v_D \sim -7$~\kms, relative to the inferred COM velocity.
However, this structure relies on the ability to distinguish lines formed in 
regions A and C separated by only 4~\kms, close to the resolution of the 
photographic spectrograms used, and moreover, lines formed at B tend to be 
weak and are not very numerous. Therefore, Willson et al.\ (1982) 
acknowledge that a single-shock structure is in principle possible as well. 
As weak lines are not sampled by the K0\,III mask used by ELODIE, the 
presence of two shocks in the atmosphere of RT\,Cyg cannot be tested with 
this mask. The tomographic method, more appropriate to investigate this 
question as described in Sect.~\ref{Sect:tomo}, provides however no evidence 
for a second shock from the mask sampling weak lines (see CCF on first line 
of Fig.~\ref{Fig:tomo}).

\subsection{X\,Oph}

Quite interestingly, X\,Oph does not exhibit the same line-doubling 
phenomenon as RT\,Cyg. This star is classified as a M5e-M9e Mira star, with 
a mean period of 328.85~days and a visual amplitude of 3.3~mag ($5.9 \le V 
\le 9.2$) according to the {\it GCVS}. Its maximum light was reached on 1999 
September, 16 (Mattei 1999, Observations from the AAVSO International 
Database, private communication). X\,Oph is a visual binary with a K1 giant 
companion (of magnitude $V = 8.9$; Merrill 1923b) in an orbit of semi-major 
axis 0.''34 and of period about 500~y (Baize \& Petit 1989). According to 
the AAVSO light curve, the faintest magnitude reached by X\,Oph during the 
survey was 7.7, so the presence of the companion did not disturb in any way 
our measurements.

The sequence of CCFs is presented in Fig.~\ref{Fig:suivi}. The CCF 
remained single during the whole monitoring (a conclusion also reached by
Hinkle et al.\ 1984 from infrared CO lines monitored over a large fraction 
of the light cycle), and the velocity stayed almost constant at 
$-$73~km~s$^{-1}$ (it varied between $-$72.8 and $-$73.4~km~s$^{-1}$). 
Nevertheless, Fig.~\ref{Fig:suivi} clearly reveals that the contrast of 
the CCF decreases with advancing phase. 

From CO millimeter observations, Groenewegen et al.\ (1999) derive for 
X\,Oph a COM velocity of $-55\pm1$~\kms\ with respect to the local standard 
of rest, or $-$73.5~\kms\ with respect to the solar barycentre. This COM 
velocity is close to the radial velocity derived from the CCF profiles: 
surprisingly, the photospheric layers probed by the K0\,III mask are almost 
at rest with respect to the COM.

\subsection{Discussion}

At this point, two important conclusions may be drawn: (i) {\it for the 
first time, clear evidences have been obtained that the \sch\ mechanism 
operates in at least some LPV stars (such as RT\,Cyg)}, (ii) {\it not all 
LPVs exhibit line doubling around maximum light though}. The second 
conclusion is confirmed by a survey of 75 LPVs observed around maximum 
light, which reveals that about 48\% of the observed stars did not exhibit 
the line doubling phenomenon with the K0\,III mask (Alvarez et al., in 
preparation). 
Nevertheless, this result does not necessarily imply that shock waves are 
not present in the deep atmosphere of those stars. Indeed, Balmer emission 
lines, which are commonly believed to be formed in the hot wake of the 
shock, are observed, although not very intense, in the spectra of X\,Oph 
(see Sect.~\ref{Sect:Balmer}). The absence of \sch\ mechanism may either be 
due to the combined effects of a weak shock and an insufficient resolving 
power to resolve the small Doppler shift between the red and blue peaks 
(that possibility seems however unlikely in the case of X\,Oph as the width 
of the CCF does not change around maximum light; see 
Fig.~\ref{Fig:suivi}), to optical-depth effects (the shock is either 
optically thin or propagates in optically-thick layers), or simply to the 
use of an inadequate template. The latter possibility will be investigated 
in a forthcoming paper of this series which will also attempt to identify 
the key stellar parameter(s) governing the appearance of the line doubling 
and/or the strength of the shock.
    
\section{Tomography}
\label{Sect:tomo}

The cross-correlation technique in the optical spectrum opens the 
possibility of much more detailed studies of pulsating atmospheres. We 
briefly describe here the method that we are currently developing to perform 
the tomography of the envelope of LPV stars. A more thorough description 
will be presented elsewhere (Alvarez et al., in preparation).
The method rests on our ability to construct reliable synthetic spectra of 
late-type giant stars, which has made significant progress in recent years 
(Plez 1999). The starting point is the `depth function' $D = D(\lambda)$, 
derived from models of late-type giant stars, and expressing the geometrical 
depth corresponding to optical depth 2/3 at the considered wavelength 
$\lambda$. The different masks $M_i$ are constructed from the collection of 
$N$ lines $\lambda_{i,j} (1 \le j \le N)$ such that 
$D_i \le D(\lambda_{i,j}) < D_i + \Delta D$, where $\Delta D$ is 
some constant optimized to keep enough lines in any given mask without 
losing too much resolution in terms of geometrical depth. Each mask $M_i$ 
should thus probe lines forming at (geometrical) depths in the range 
$D_i, D_i + \Delta D$ in the atmosphere. 
These masks are then used as templates to correlate with the observed 
spectra of the LPV stars. This procedure should provide the velocity field 
as a function of depth in the atmosphere of the LPV stars.  
One crucial requirement  of the method is that the masks constructed from 
static atmospheric models do not lose their ability to probe a layer of a 
{\it given} (geometrical) depth when applied to a dynamic atmosphere. This
point will be discussed in more details in a forthcoming paper.

\begin{figure}[h]
  \resizebox{\hsize}{!}{\rotatebox{90}{\includegraphics{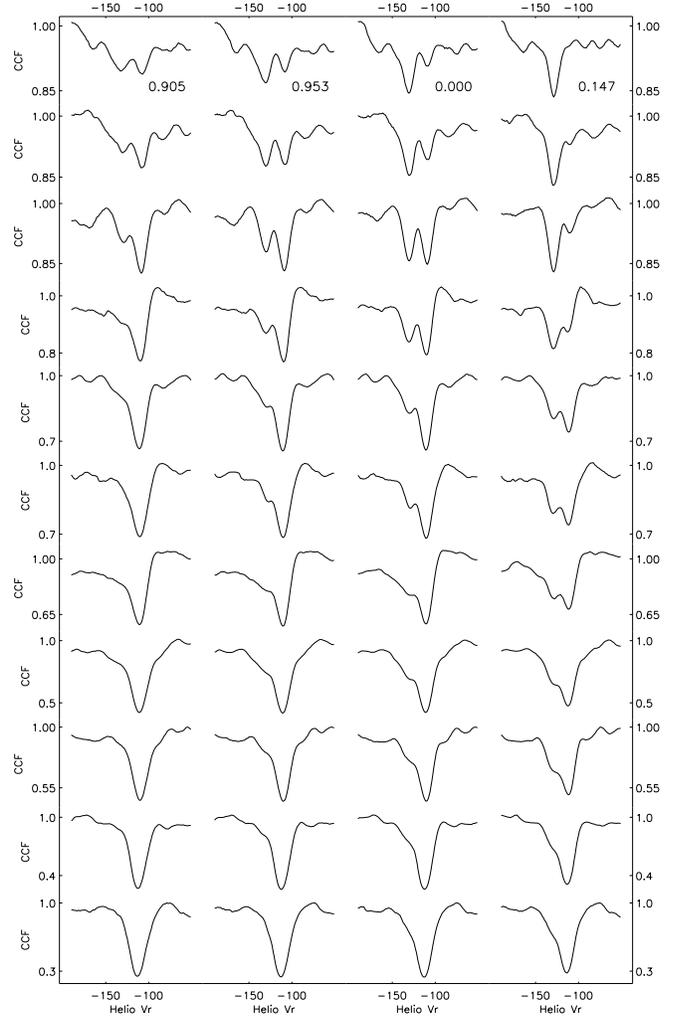}}}
  \caption[]{Sequence of CCFs probing layers of increasing depth (from 
  bottom to top) in the atmosphere of RT\,Cyg at 4 phases (0.905, 0.953, 
  0.000 and 0.147)}
  \label{Fig:tomo}
\end{figure}

Fig.~\ref{Fig:tomo} displays a sequence of CCFs probing increasing depths 
in the atmosphere of RT\,Cyg at four phases. At phase 0.905, it can clearly 
be seen that the line doubling occurs only for the 3 deepest masks. At a 
later phase (0.953), the line doubling involves 3 more masks further out, 
translating the upward motion of the shock. At even later phases (0.000 and 
0.147), the line doubling is seen up to the outermost mask. At phase 0.147,
it is worth noting that the deepest mask displays a {\it single blue} peak, 
while the outermost mask displays a single red peak (though somewhat 
asymmetrical). In summary, the \sch\ scenario can be observed on both the
temporal {\it and} spatial variations of the CCFs displayed in 
Fig.~\ref{Fig:tomo}, thus clearly revealing the upward motion of the shock
wave.

\section{The Balmer lines}
\label{Sect:Balmer}

The Balmer lines are important diagnostics to which the temporal behaviour 
of the CCF of absorption lines may be compared. It is not the purpose of 
this paper to discuss thoroughly the different theories concerning the 
origin of the emission lines. Nevertheless, thanks to the two-month-long 
monitoring, the possibility is offered to present a high-quality temporal 
sequence of Balmer lines, which might later on be used to constrain the 
different existing models.

\subsection{RT\,Cyg}

\begin{figure*}
  \resizebox{\hsize}{!}{\rotatebox{90}{\includegraphics{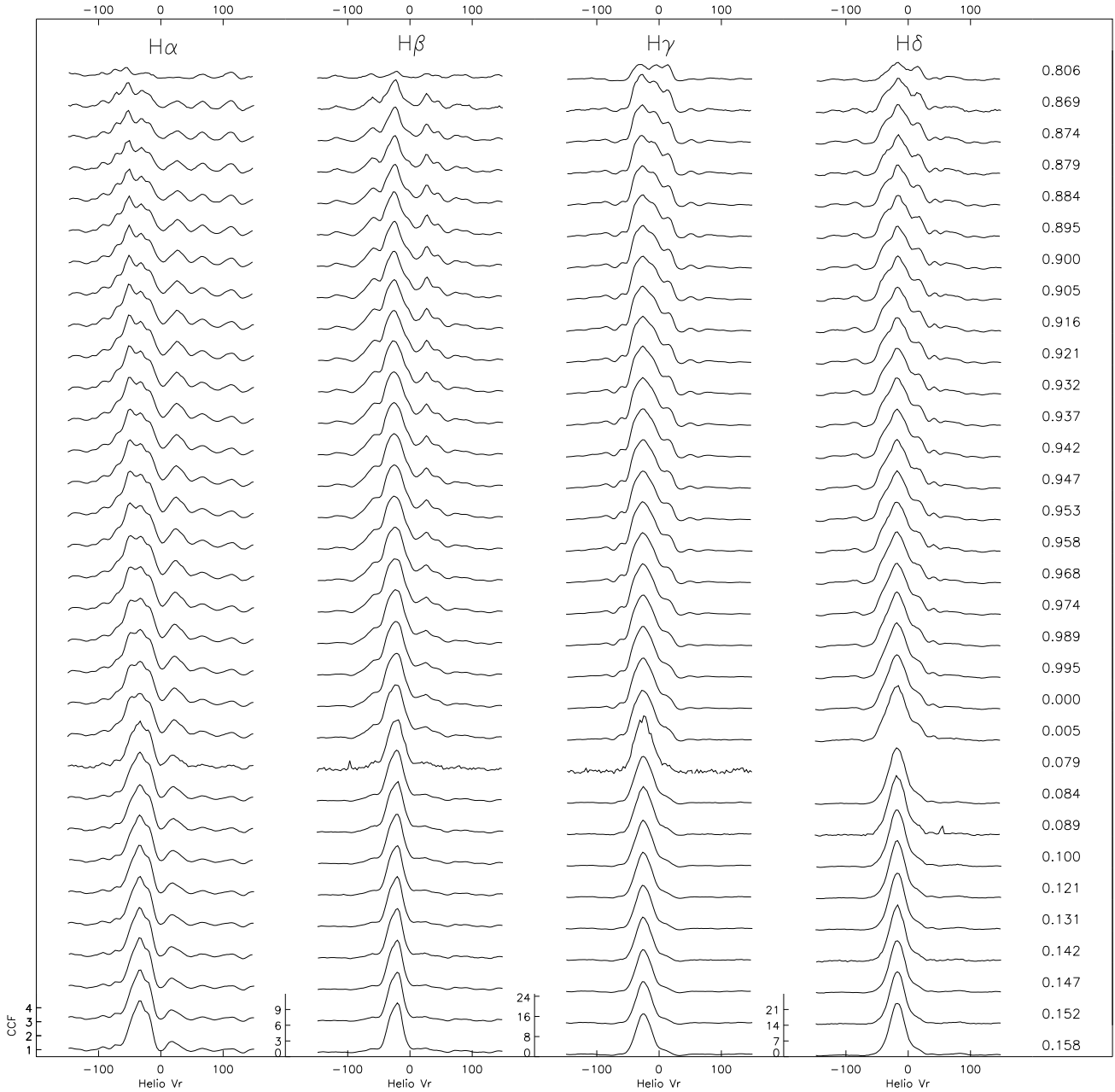}}}
  \caption[]{Evolution of the \ha, \hb, \hg\ and \hd\ profiles of RT\,Cyg
  from phases 0.806 to 1.158. The vertical axes (only indicated for the sake 
  of clarity for the last profile of each sequence) have been normalized to 
  the average value of the continuum in the same spectral order, away from 
  the emission line(s). The abcissa scale is a velocity scale, with its zero 
  point adjusted to the average velocity of the red component of the 
  absorption lines (i.e.\ $-$110~km~s$^{-1}$; see text)}
  \label{Fig:rtcyg_balmer}
\end{figure*}

Figure~\ref{Fig:rtcyg_balmer} displays the evolution of the \ha, \hb, \hg\ 
and \hd\ profiles of RT\,Cyg during the monitoring. The abcissa scale is a 
velocity scale. Since the COM velocity is not known for RT\,Cyg, the zero 
point of the velocity scale has been taken equal to a heliocentric velocity 
of $-$110~km~s$^{-1}$ (corresponding to the mean velocity of the red 
component of the CCF profiles; see Fig.~\ref{Fig:Vr_rtcyg}). When analyzing 
Fig.~\ref{Fig:rtcyg_balmer}, it should thus be remembered that the zero 
point of the velocity scale has been adjusted so as to correspond to the 
velocity of the red component of the absorption lines, which is almost 
certainly different from the stellar COM velocity. The \hd\ profile is not 
shown for phase 0.079 because of the too low signal-to-noise ratio achieved 
in this order.

The \ha\ profile is seen in emission during the whole monitoring, although
weakly so on the first frame corresponding to phase 0.806. In the early
development of the emission feature, it is obliterated by an absorption 
feature falling right on top of it. At later phases, a separate red 
component is very clearly present. Such a structure has been observed in 
several Mira stars (see Gillet 1988a; Woodsworth 1995). The \ha\ and \hb\ 
profiles are rather similar in the first half of the monitoring, but, at the 
latest phases, the \hb\ profile shows no red component but instead a very 
clear pedestal at the base of the line. The \hg\ and \hd\ emissions are much 
more complex in the early stages of the monitoring. Mutilations of the \hg\ 
and \hd\ profiles are expected to be less pronounced than in the case of 
\ha\ and \hb\, as molecular absorptions are weaker in their spectral domain 
(Gillet 1988b).

\subsection{X\,Oph}

\begin{figure*}
  \resizebox{\hsize}{!}{\rotatebox{90}{\includegraphics{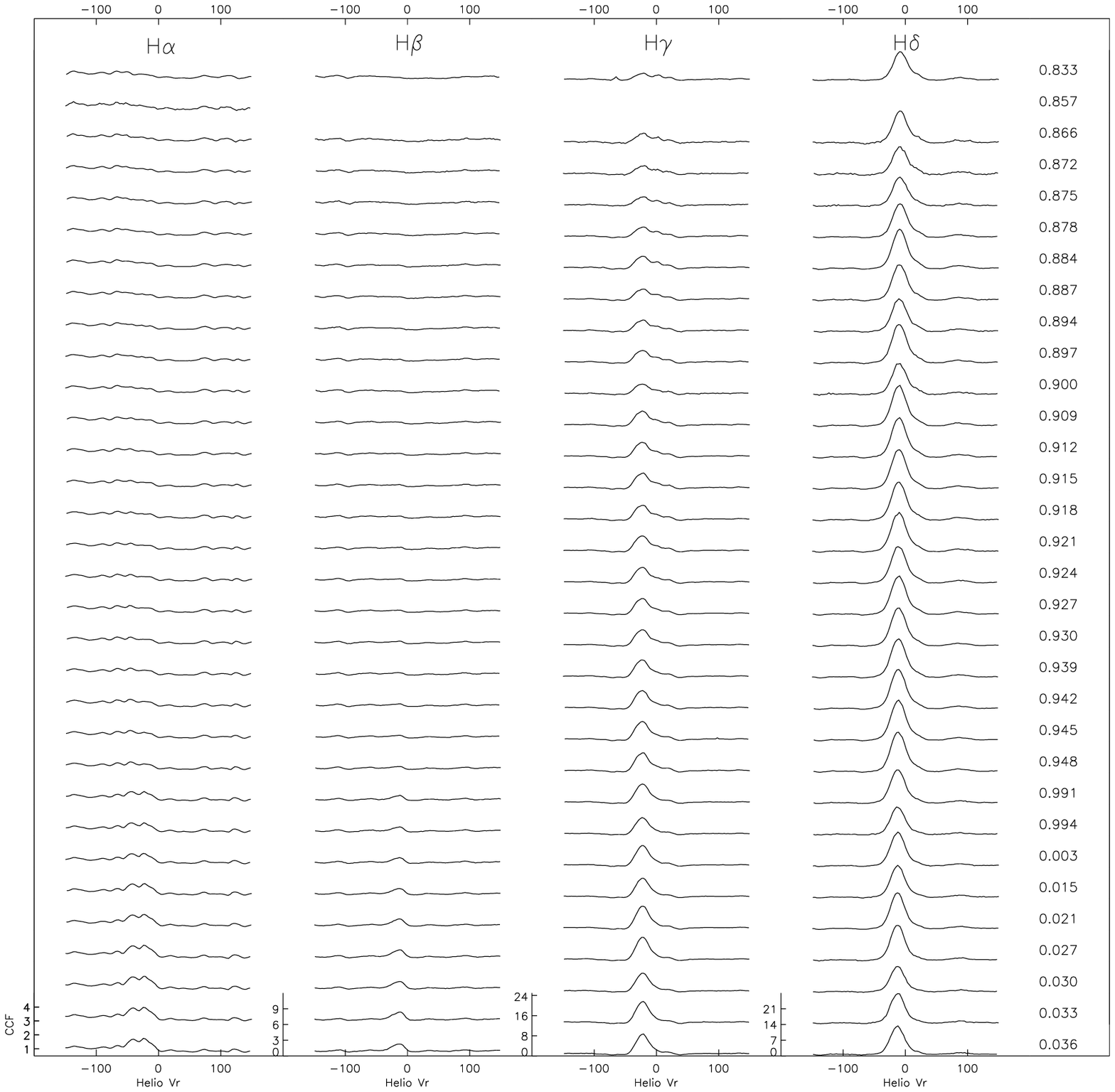}}}
  \caption[]{Evolution of the \ha, \hb, \hg\ and \hd\ profiles of X\,Oph
  from phases 0.833 to 1.036. The vertical axes have been normalized to 
  the average value of the continuum in the same spectral order, away from 
  the emission line(s). The abcissa scale is a velocity scale, with its 
  zero point adjusted to the COM velocity as derived by Groenewegen et al.\ 
  (1999)}
  \label{Fig:xoph_balmer}
\end{figure*}

Figure~\ref{Fig:xoph_balmer} displays the temporal evolution of the Balmer 
lines of X\,Oph during the monitoring. The COM velocity has been taken equal 
to $-$73.5~\kms\ (see Sect.~\ref{Sect:monitoring}.3). The vertical-axis scale 
is the same as in the case of RT\,Cyg to make the comparison with 
Fig.~\ref{Fig:rtcyg_balmer} easier. The \hb, \hg\ and \hd\ profiles are not 
shown for phase 0.857 because of the too low signal-to-noise ratio achieved 
in the blue orders.

Although \hd\ is almost as strong in X\,Oph as in RT\,Cyg (but not quite 
so), \hg\ is much weaker (by at least of factor of 2), and \hb\ and \ha\ are 
only barely visible after phase 0.991. It is interesting to note that the 
relative intensities of the Balmer lines are similar in X\,Oph and RT\,Cyg. 
Therefore, it is not surprising that \ha\ never gets very strong in X\,Oph, 
since a look at RT\,Cyg shows that, for \ha\ to be clearly visible, the 
intensity of \hd\ needs to exceed 20 (relative to the local continuum) which 
is never the case in X\,Oph. This may indicate that the same 
(recombination?) mechanism is operating in both stars to produce the Balmer 
emission lines and that the variation of the underlying opacity with 
wavelength is the same as well.

The absence of \sch\ mechanism in X\,Oph and its occurrence in RT\,Cyg may 
similarly be related  to  their respective Balmer line intensities. The CCF 
profile starts becoming clearly asymmetric at phase 0.884 in RT\,Cyg 
(Fig.~\ref{Fig:suivi}), when the relative intensity of \ha\ amounts 
to 3. As such intensity levels are never reached in X\,Oph, it is thus not 
surprising that the CCF remains single, provided of course that the \sch\ 
mechanism and the Balmer line intensities are indeed somehow correlated. 
Whether or not this is systematically the case remains however to be 
confirmed on a larger sample. At this point, it may just be concluded that 
RT\,Cyg and X\,Oph differ by the presence/absence of the \sch\ mechanism
{\it and} by the strength of the Balmer lines. 

\subsection{Comparing the velocities of the Balmer and absorption lines 
in RT\,Cyg}

The comparison of the \ha\ and absorption-line velocities illustrates the 
respective merits of the competing models accounting for the complex \ha\
line profile.
Two different models involving a single spherical shock wave have been 
proposed to explain the origin of the double \ha\ emission structure 
observed in Mira stars  (Gillet 1988b). The first one is the so-called 
`geometrical model' (Willson 1976; Gillet et al.\ 1985) stating that, while 
the emission line is produced within the de-excitation zone behind the shock 
front, the redshifted component is emitted by the shock propagating far 
above the photosphere in the hemisphere opposite to the observer whereas the 
blueshifted component is emitted by the shock propagating in the hemisphere 
facing the observer. In that model, the two emission components should be 
almost centered on the stellar COM velocity. 
The second interpretation is the self-reversal model (Bidelman \& Ratcliffe 
1954; Gillet 1988a) stating that there is a true absorption obliterating the 
emission line, caused by the presence of cool hydrogen gas above the shock 
front. In the latter interpretation, the central absorption should have the 
velocity of the layers lying above the shock. In the framework of the \sch\ 
model, those layers are responsible for the red absorption-line component. 
As the velocity of those layers approximately corresponds to the adopted 
zero-point of the velocity scale for the Balmer features, the central 
absorption should  have a velocity close to zero, which indeed appears to be 
the case, as indicated below in relation with Fig.~\ref{Fig:velbalmer}.  

\begin{figure}
  \resizebox{\hsize}{!}{\includegraphics{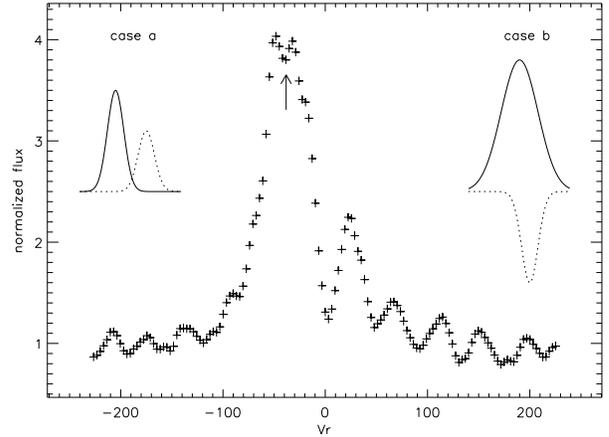}}
  \caption[]{Example of \ha\ profile in RT\,Cyg (phase 0.953). Gaussian fits
  have been made according to {\it case a} (two emission lines) or 
  {\it case b} (broad emission and central absorption). The arrow indicates 
  an absorption feature (see text)}
  \label{Fig:RTCyg_Ha_mod}
\end{figure}

\begin{figure}
  \resizebox{\hsize}{!}{\includegraphics{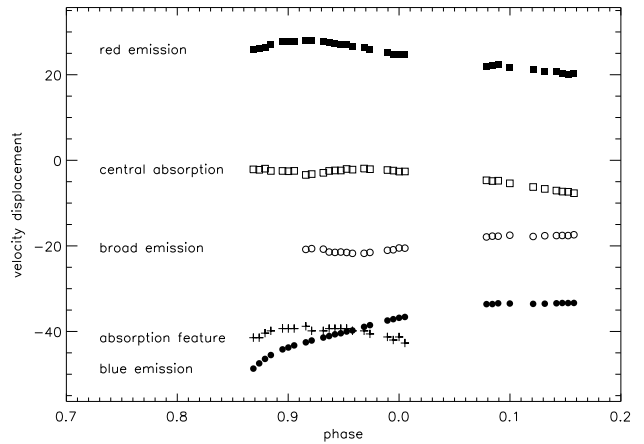}}
  \caption[]{Evolution with phase of the velocity structure of \ha\ in 
  RT\,Cyg according to two models (see text):
  (i) geometric model ({\it case a}): filled circles: blue emission; filled 
  squares: red emission,
  (ii) self-reversal model ({\it case b}): open circles: broad emission; 
  open squares: central absorption.
  Crosses: atomic or molecular absorption feature unrelated to \ha}
  \label{Fig:velbalmer}
\end{figure}

An example of \ha\ profile obtained during the monitoring (phase 0.953) of
RT\,Cyg is shown in Fig.~\ref{Fig:RTCyg_Ha_mod}. To evaluate the merits of 
the above two models\footnote{It is worth mentioning that a third 
interpretation, not considered here, has been proposed by Woodsworth (1995) 
and involves three emission components.}, the complex line profile has been 
fitted by gaussian functions in two different ways. The `geometrical model' 
requires to fit separately the blue and red emission components 
({\it case a} on Fig.~\ref{Fig:RTCyg_Ha_mod}) whereas the self-reversal 
model requires to fit separately the broad emission line and the central 
absorption ({\it case b} on Fig.~\ref{Fig:RTCyg_Ha_mod}).
Figure~\ref{Fig:velbalmer} displays the velocity structure of the \ha\ line 
as a function of phase, for the two cases. It should also be remembered that 
there is an atomic or molecular absorption feature unrelated to \ha\ falling
right on top of it, as indicated by the arrow in 
Fig.~\ref{Fig:RTCyg_Ha_mod}. Its velocity is also displayed in 
Fig.~\ref{Fig:velbalmer} for phases earlier than 0.01, since it is no more 
visible afterwards. Because these absorption features mutilate the still 
weak \ha\ line during the early stages of the monitoring, the fit of the 
broad emission feature performed for {\it case b} is rather inaccurate and 
has therefore not been presented on Fig.~\ref{Fig:velbalmer} until phase 
0.91 when  the fit becomes more reliable. 

Figure~\ref{Fig:velbalmer} offers the interesting possibility to compare the
evolution of the structure of the \ha\ line with the corresponding evolution 
of the absorption features (Fig.~\ref{Fig:Vr_rtcyg}). The comparison of 
these two figures reveals interesting similarities in the framework of 
{\it case b}. First, the velocity of the central absorption feature 
overlying the broad \ha\ line has a velocity similar to that of the red 
component of the absorption lines, as expected if both features form in the 
layers just above the shock. Second, the velocity difference between the 
broad emission line and the overlying absorption feature 
($\sim 10 - 20$~\kms) is of the same order as the difference between the 
velocities of the red and blue peaks of the absorption lines. This is again 
consistent with the hypothesis that the broad \ha\ emission and the blue 
absorption lines both form in the ascending, post-shock layers. 

On the contrary, there is no clear correspondence between the 
absorption-line velocities and the velocities of the red and blue \ha\ 
emission peaks defined in the framework of {\it case a}. 

At this point, one should mention that the interpretation of the \ha\ line 
profile according to {\it case a} or {\it b} also results in very different 
shock-front velocities (Gillet 1988a). The geometrical model ({\it case a}) 
yields rather large shock velocities (of the order of 60~\kms), whereas the
self-reversal model ({\it case b}) involves much smaller shock velocities 
($<$30~\kms; Gillet 1988a). On theoretical grounds, Gillet et al.\ (1989b)
have shown that a shock velocity in excess of 60~\kms\ is required for the 
shock to be strong enough to first photodissociate H$_2$ molecules and then 
photoionize the resulting hydrogen atoms, to produce the observed Balmer
emission lines by recombination. 
This requirement is consistent with the velocities (50--70~\kms) deduced 
from the observations of fluorescent lines in Miras (Willson 1976), and 
would also favour {\it case a} over {\it case~b}.

In the above discussion, {\it case a} faced however the difficulty that 
there is no clear coincidence between the \ha\ and absorption-line 
velocities, although one would naively expect some if those lines are formed 
in the vicinity of the same shock. The above conclusion of discrepant 
velocities has been reached assuming that both \ha\ and absorption lines 
form in a thin layer, so that they can be assigned the velocity of the layer 
where they form. However, this assumption might not hold true for absorption 
lines which form over a broader region than \ha, so that the velocity of the 
absorption lines result from an average of the velocity field over the 
line-forming region. The discrepant \ha\ and absorption-line velocities 
observed in the framework of {\it case a} may possibly be reconciled when 
such averaging effects are properly taken into account.

Detailed dynamic models coupled to radiative transfer are thus ultimately 
needed to answer the question of the origin of the complex \ha\ line 
profile.

\section{Conclusion}

The origin of double-peaked emission and absorption lines in the complex LPV 
atmospheres is a long-standing problem. If the shock wave and the associated 
velocity gradient are generally believed to be at the origin of the complex 
line profiles of LPV stars (de la Reza 1986; Gillet 1988b; Hinkle et al.\ 
1997), alternative models not resorting to differential atmospheric motions 
are sometimes invoked. They involve either purely non-LTE effects to account 
for the double-peaked Balmer emission (Magnan \& de Laverny 1997), or a 
temperature inversion to account for the splitting of the absorption lines 
(Karp 1975; Gillet et al.\ 1985).

Thanks to the cross-correlation technique and a daily survey of the Mira 
star RT\,Cyg, the role of velocity fields in the formation of genuinely 
double absorption lines has been undoubtedly confirmed in the framework of 
the \sch\ mechanism. Evidence for the existence of strong shock waves in the 
outer layers of LPVs becomes compelling, although not all Mira stars exhibit 
the \sch\ scenario, as demonstrated by the observations of X\,Oph collected 
in the present paper. The origin of this distinct behaviour will be 
investigated in the next paper of this series.

A tomographic technique that opens new perspectives for the study of the 
dynamics of LPVs has been used to visualize the outward motion of the shock 
wave in the atmosphere of RT\,Cyg. 

The evolution of the \ha, \hb, \hg\ and \hd\ emission-line profiles during 
35\% (20\%) of the cycle of RT\,Cyg (X\,Oph) around maximum light has also 
been discussed. The comparison of this data set with the temporal evolution 
of the profile of absorption lines offers interesting perspectives to 
identify the origin of the complex profile of the \ha\ emission line. 

The present paper had no other ambition than providing a description of the 
data collected with the ELODIE spectrograph around maximum light for RT\,Cyg 
and X\,Oph. These data may be used as constraints for future detailed 
models. Firm conclusions on the line-formation mechanisms in Mira variables 
should thus await the advent of such detailed dynamic models coupling 
hydrodynamics to radiative transfer.

\begin{acknowledgements}
We are grateful to X.\ Delfosse, D.\ Erspamer, A.\ Gomez, M.\ Haywood, 
E.\ Josselin, M.\ Mayor, D.\ Naef, D.\ Segransan, E.\ Oblak and S.\ Udry for
performing part of the observations, as well as to the staff of the 
Haute-Provence Observatory. 
Ephemeris and light curves provided by the AAVSO International Database were 
very helpful for planning the monitoring described in the present paper. 
R.A.\ benefits of a TMR ``Marie Curie'' Fellowship at ULB. A.J.\ is Research 
Associate from the {\it Fonds National de la Recherche Scientifique} 
(Belgium).
\end{acknowledgements}

\end{document}